\documentclass[twocolumn,aps,prd,preprintnumbers,nofootinbib]{revtex4-2}
\usepackage{graphicx}
\usepackage{amsmath}
\usepackage{amssymb}
\usepackage{amsfonts}
\usepackage{hyperref}
\usepackage{url}
\usepackage{color}
\usepackage{bm}

\newcommand{\be}{\begin{equation}}
\newcommand{\ee}{\end{equation}}
\newcommand{\ba}{\begin{eqnarray}}
\newcommand{\ea}{\end{eqnarray}}
\newcommand{\nn}{\nonumber}

\renewcommand{\[}{\begin{equation}}
\renewcommand{\]}{\end{equation}}

\usepackage{array}
\makeatother

\begin{document}

\preprint{IFT-UAM/CSIC-21-67}

\title{Machine Learning improved fits of the sound horizon at the baryon drag epoch}

\author{Andoni Aizpuru}
\email{andoni.aizpuru@estudiante.uam.es}

\author{Rub\'{e}n Arjona}
\email{ruben.arjona@uam.es}

\author{Savvas Nesseris}
\email{savvas.nesseris@csic.es}

\affiliation{Instituto de F\'isica Te\'orica UAM-CSIC, Universidad Auton\'oma de Madrid,
Cantoblanco, 28049 Madrid, Spain}

\date{\today}

\begin{abstract}
The baryon acoustic oscillations (BAO) have proven to be an invaluable tool in constraining the expansion history of the Universe at late times and are characterized by the comoving sound horizon at the baryon drag epoch $r_\mathrm{s}(z_\mathrm{d})$. The latter quantity can be calculated either numerically using recombination codes or via fitting functions, such as the one by Eisenstein and Hu (EH), made via grids of parameters of the recombination history. Here we quantify the accuracy of these expressions and show that they can strongly bias the derived constraints on the cosmological parameters using BAO data. Then, using a machine learning approach, called the genetic algorithms, we proceed to derive new analytic expressions for $r_\mathrm{s}(z_\mathrm{d})$ which are accurate at the $\sim0.003\%$ level in a range of $10\sigma$ around the Planck 2018 best-fit or $\sim0.018\%$ in a much broader range, compared to $\sim 2-4\%$ for the EH expression, thus obtaining an improvement of two to three orders of magnitude. Moreover, we also provide fits that include the effects of massive neutrinos and an extension to the concordance cosmological model assuming variations of the fine structure constant. Finally, we note that our expressions can be used to ease the computational cost required to compute $r_\mathrm{s}(z_\mathrm{d})$ with a  Boltzmann code when deriving cosmological constraints using BAO data from current and upcoming surveys.
\end{abstract}

\maketitle

% -------------------
\section{Introduction \label{sec:introduction}}
Some of the strongest constraints on the expansion of the Universe at late times come from baryon acoustic oscillations (BAO) data. The BAO were formed in the early Universe, while it was very homogeneous (as probed today by the CMB) except for tiny fluctuations, and the  photons and baryons were tightly coupled \cite{dodelson2003modern}. As the Universe expanded, it became cooler and less dense, while the fluctuations grew due to gravity. Acoustic waves were generated as the photon-baryon fluid was attracted and fell onto the overdensities producing compressions and rarefactions due to the gravitational collapse and radiation pressure. 

These acoustic waves propagated until the Universe became cool enough for the electrons and protons to recombine and then the baryons and photons decoupled. The time when the baryons were “released” from the drag of the photons is known as the drag epoch, $z_\mathrm{d}$ \cite{weinberg2008cosmology}. From then on, photons expanded freely while the acoustic waves “freezed in” the baryons in a scale given by the size of the sound horizon at the drag epoch, dubbed $r_\mathrm{s}(z_\mathrm{d})$. Progressively, baryons fell into dark matter potential wells but also dark matter was attracted to baryon overdensities. Neutrinos did not interact, so they streamed away while dark matter responded to gravity and fell onto the overdensity. 

The perturbations were dominated by photons and baryons as they were coupled, resulting in overdensities and overpressure which tried to equalize with the surrounding resulting in an expanding sound wave moving at the speed of sound, approximately $c_s^2\sim 1/3$. The perturbation in photons and baryons was carried outward and the photons and baryons continued to expand whereas neutrinos spread out. Dark matter continued to fall into perturbations, which kept growing. 

As the expanding Universe continued to cool down, it reached a point when the electrons and protons began to combine. Since photons did not scatter as efficiently they started to decouple. The sound speed dropped and the pressure wave slowed down. The process continued until the photons where completely decoupled and then the perturbations smoothed out\footnote{\href{http://mwhite.berkeley.edu/BAO}{http://mwhite.berkeley.edu/BAO}}. In fact, the sound speed of the baryon perturbation dropped so much that the pressure wave stalled. Thus, the original dark matter  perturbation was left surrounded by a baryon perturbation in a shell. The two components attracted each other and the perturbations started to mix\footnote{\href{https://lweb.cfa.harvard.edu/~deisenst/acousticpeak/}{https://lweb.cfa.harvard.edu/$\mathrm{\sim}$deisenst/acousticpeak/}}. 

The BAO provides a characteristic scale that is “frozen” in the galaxy distribution providing a standard ruler that can be measured as a function of redshift in either the galaxy correlation function or the galaxy power spectrum. The BAO determination of the  geometry of the Universe is quite robust against systematics and has been measured by several surveys, such as  the SDSS \cite{deMattia:2020fkb} and 2dFGRS \cite{Percival:2007yw}. The BAO signature provides a standard ruler that can be used to measure the geometry of the Universe and it can measure both the angular diameter distance $d_A(z)$ and the expansion rate $H(z)$. Measurements of the BAO only provide the combination of $H_0$ and $r_\mathrm{s}(z_\mathrm{d})$, which means the two parameters are fully degenerate.  As a result, the constraints obtained from the analysis of the BAO can be influenced significantly on the assumption of $r_\mathrm{s}(z_\mathrm{d})$ \cite{Cuceu:2019for}.

In order to accurately estimate $r_\mathrm{s}(z_\mathrm{d})$, one may use either recombination codes, such as \texttt{RECFAST} \cite{Seager:1999bc}, \texttt{CosmoRec} \cite{Chluba:2010ca} or \texttt{HyRec} \cite{Lee:2020obi,AliHaimoud:2010dx}, or analytic approximations based on fits of grids of parameters of the recombination history. A prominent example of the latter approach is the formula by Eisenstein and Hu \cite{Eisenstein:1997ik}, hereafter known as EH, which provides a fit of  $r_\mathrm{s}(z_\mathrm{d})$ in terms of the matter and baryon density parameters. This formula has been extensively used in the literature in analyses of the BAO data, see for example Refs.~\cite{Beutler:2011hx,Komatsu:2010fb,Bamba:2012cp,Zhai:2018vmm,
Martinelli:2020hud}. However, as already observed in Ref.~\cite{Eisenstein:1997ik}, this expression is only accurate to the $\sim2\%$ level and as a result is not appropriate for deriving cosmological constraints from BAO data in a percent cosmology era with current and upcoming surveys.

Over the years attempts to improve the EH formula have appeared. For example, the dependence of $r_\mathrm{s}(z_\mathrm{d})$ on various parameters, including massive and massless neutrinos, was examined in Ref.~\cite{Thepsuriya:2014zda}. On the other hand, fits of $r_\mathrm{s}(z_\mathrm{d})$ including neutrinos and relativistic species were found in Ref.~\cite{Aubourg:2014yra} and in Ref.~\cite{Anderson:2013zyy}. Finally, how the fraction of the baryonic mass in Helium $Y_P$ and the relativistic degrees of freedom $N_\mathrm{eff}$ affects the sound horizon and how both are degenerate, was studied in Ref.~\cite{Hou:2011ec}.

The main limitation of the aforementioned analyses is that some ad-hoc parametrizations were fitted to grids of parameters and $r_\mathrm{s}(z_\mathrm{d})$, thus being limited from the start on how accurate they can be. Hence, in our work we use machine learning to provide, in a data driven approach, extremely accurate fits to the comoving sound horizon at the baryon drag epoch $r_\mathrm{s}(z_\mathrm{d})$. We then compare these expressions against both the original formula of EH and the exact numerical estimation of the sound horizon, in order to quantify the amount of bias this expression introduces in the constraints.

In our analysis we also consider separately the effect of massive neutrinos and a varying fine structure constant and we find that our fits provide an improvement of a factor of three compared to other simple parametrizations and can be used in current and upcoming surveys to derive cosmological constraints so as to ease the computational cost that would be required when computing $r_\mathrm{s}(z_\mathrm{d})$ via a  Boltzmann code.

The structure of our paper is as follows: in Sec.~\ref{sec:analysis} we present the theoretical background and main assumptions in our work, while in Sec.~\ref{sec:GA} we present some details on our machine learning approach used to improve the sound horizon fits. In Sec.~\ref{sec:results} we present our main results, while in Sec.~\ref{sec:conclusions} we summarize our conclusions. Finally, in Appendix~\ref{sec:zd} we present some complementary fits for the redshift at the drag and recombination epochs.

\section{Theory\label{sec:analysis}}

The comoving sound horizon at the drag epoch is given by
\begin{equation}\label{eq:rs_eq}
r_{\mathrm{s}}\left(z_{\mathrm{d}}\right)=\frac{1}{H_{0}} \int_{z_{\mathrm{d}}}^{\infty} \frac{c_{\mathrm{s}}(z)}{H(z) / H_{0}} \mathrm{~d} z,
\end{equation}
where $z_{\rm d}$ is the redshift at the drag epoch, see Eq.~(4) of Ref.~\cite{Eisenstein:1997ik}, while $c_{\rm s}(z)$ is the sound speed in the baryon-photon fluid given by
\be
c_{\rm s}=\frac{c}{\sqrt{3(1+R)}},
\ee 
where $R=\frac{3 \rho_{b}}{4 \rho_{\gamma}}=\frac{3 \Omega_{b,0}}{4 \Omega_{\gamma,0}}a$ and $c$ is the speed of light in vacuum. By definition, the sound horizon at the baryon drag epoch is the comoving distance a wave can travel prior to $z_\mathrm{d}$ and it depends on the epoch of recombination, the expansion of the Universe and the baryon-to-photon ratio. The sound horizon is well determined by the Cosmic Microwave Background (CMB) measurements of the acoustic peaks.

Regarding the neutrinos, neutrino flavour oscillation experiments have shown that they are massive \cite{bilenky2016neutrino}, providing a direct evidence for physics beyond the Standard Model. Cosmology is a very propitious stage to probe neutrino properties since they leave an imprint in the CMB and in the distribution of Large-Scale Structure (LSS) in the Universe. The energy density of massive neutrinos, $\rho_{\nu}=\sum m_{\nu,i}n_{\nu,i}$, corresponds to
\be
\Omega_{\nu}h^2 \sim \frac{\sum m_{\nu,i}}{94\text{eV}},
\ee
where $n_{\nu}$ represents number density of neutrinos.

We also consider variations of fundamental constants, which are usually assumed to be constant over space-time. These constants are defined operationally, meaning that nature by itself does not force it to be constant. They have to be obtained experimentally since they are not given by the theory, see for instance Ref.~\cite{Landau:2020vkr} for a review on the variation of fundamental constants. Here we will examine the interesting case where the fine structure constant, defined in laboratory scales at late times as $\alpha_{0}=\frac{e^2}{\hbar c}$, is rescaled and we will express its relative variation over its standard model value as $\alpha/\alpha_{0}$. Thus, we assume that $\alpha$ is the value at early times of the fine structure constant and is rescaled with respect to its laboratory (late time) value $\alpha_{0}$, with a sharp transition at intermediate redshifts.

If there are eventually signatures of a variation it would have imprints in different physical mechanisms such as the CMB anisotropies \cite{Uzan:2010pm}. Constraints on this variation, both temporal and spatial, have been performed already \cite{Clara:2020efx,deMartino:2016bjx,deMartino:2016tbu,Hees:2014lfa,Colaco:2020ndf,Lopez-Honorez:2020lno,Wilczynska:2020rxx}, and this variation can be produced for example through an evolving scalar field which is coupled to the electromagnetic Lagrangian \cite{Taylor, Casas1, Casas2, deMartino:2016bjx}. This will give rise to variations of the fine structure constant, a violation of the Weak Equivalence principle and violations of the standard $T_\mathrm{CMB}(z)$ law, as the number of photons is no longer conserved. These kinds of models can in principle be constrained by future large scale structure surveys using high-resolution spectroscopic data in combination with local astrophysical data, see
Ref.~\cite{Hart:2019dxi} for updated constraints with current data and 
Ref.~\cite{Martinelli:2021tex} for recent forecasts with upcoming surveys.

Another class of models where this occurs is the Bekenstein-Sanvik-Barrow-Magueijo (BSBM) model \cite{Sandvik:2001rv}, where the electric charge is allowed to vary. Although such theories preserve the local gauge and Lorentz invariance, the fine structure constant will vary during the matter dominated era.

\section{The Genetic Algorithms \label{sec:GA}}

In this section we will describe the Genetic Algorithms (GA) that will be used in our analysis to improve the sound horizon fits. The GA have been successfully used in cosmology for several reconstructions on a wide range of data, see for example Refs.~\cite{Bogdanos:2009ib,Nesseris:2010ep, Nesseris:2012tt,Nesseris:2013bia,Sapone:2014nna,Arjona:2020doi,Arjona:2020kco,Arjona:2019fwb,Arjona:2021hmg,Arjona:2020skf,Arjona:2020axn,Arjona:2021zac}. Other applications of the GA cover other areas such as particle physics \cite{Abel:2018ekz,Allanach:2004my,Akrami:2009hp} and astronomy and astrophysics \cite{wahde2001determination,Rajpaul:2012wu,Ho:2019zap}. Other symbolic regression methods implemented in physics and cosmology can be found at \cite{Udrescu:2019mnk,Setyawati:2019xzw,vaddireddy2019feature,Liao:2019qoc,Belgacem:2019zzu,Li:2019kdj,Bernardini:2019bmd,Gomez-Valent:2019lny}.

\begin{table}[!t]
\caption{The grammars used in the GA analysis. Other complex forms are automatically produced by the mutation and crossover operations as described in the text.\label{tab:grammars}}
\begin{centering}
\begin{tabular}{cc}
 Grammar type & Functions \\ \hline
Polynomials & $c$, $x$, $1+x$ \\
Fractions & $\frac{x}{1+x}$\\
Trigonometric & $\sin(x)$, $\cos(x)$, $\tan(x)$\\
Exponentials & $e^x$, $x^x$, $(1+x)^{1+x}$ \\
Logarithms & $\ln(x)$, $\ln(1+x)$
\end{tabular}
\par
\end{centering}
\end{table}

The GA can be regarded as a machine learning (ML) technique constructed to carry out unsupervised regression of data, i.e. it performs non-parametric reconstructions that finds an analytic function of one or more variables (like in our case here) that describes the data extremely well. The GA emulates the concept of biological evolution through the principle of natural selection, as brought by the genetic operations of mutation and crossover. 

\begin{figure}[!t]
\centering
\includegraphics[width = 0.45\textwidth]{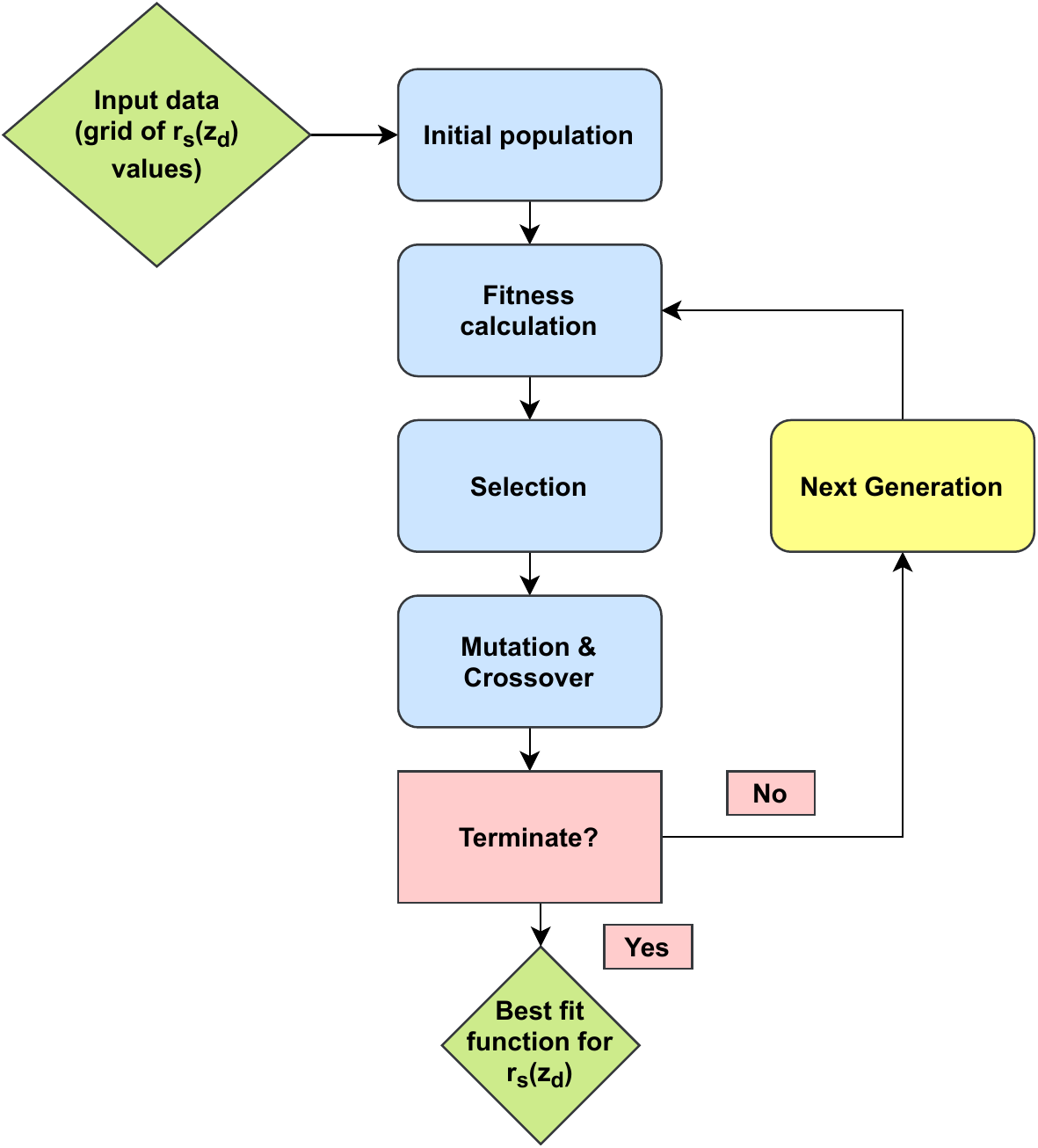}\caption{A flowchart of the list of the steps for the Genetic Algorithm reconstruction of $r_\mathrm{s}(z_\mathrm{d})$.\label{fig:flowchart}}
%\vspace*{-1mm}
\end{figure}

In essence, a set of trial functions evolves as time passes by through the effect of the stochastic operators of crossover, i.e. the joining of two or more candidate functions to form another one, and mutation, i.e. a random alteration of a candidate function. This process is then repeated thousands of times with different random seeds to ensure convergence and explore properly the functional space. In Fig.~\ref{fig:flowchart} we present a flowchart of the steps the GA goes through when reconstructing a function.

Since the GA is constructed as a stochastic approach, the probability that a population of functions will bring about offspring is principally assumed to be proportional to its fitness to the data, where in our analysis is given by a $\chi^2$ statistic and give the information on how good every individual agrees with the data. For the simulated data in our analysis we are assuming that the likelihoods are sufficiently Gaussian that we use the $\chi^2$ in our GA approach. Then, the probability to have offspring and the fitness of each individual is proportional to the likelihood causing an “evolutionary” pressure that favors the best-fitting functions in every population, hence directing the fit towards the minimum in a few generations.

In our analysis we reconstruct the $r_\mathrm{s}(z_\mathrm{d})$ function considering that it depends on the following variables: $\{\Omega_mh^2,\Omega_bh^2\}$, $\{\Omega_mh^2,\Omega_bh^2,\Omega_{\nu}h^2\}$ and $\{\Omega_mh^2,\Omega_bh^2,\alpha/\alpha_0\}$ respectively. To calculate the sound horizon we use the code \texttt{CLASS} by Ref.~\cite{Blas:2011rf} and the \texttt{HYREC-2} recombination module \texttt{Hyrec2020} \cite{Lee:2020obi,AliHaimoud:2010dx}. We then make grids of parameters and $r_\mathrm{s}(z_\mathrm{d})$ and fit the values with the genetic algorithms. For example, when $\{\Omega_mh^2=0.13,\Omega_bh^2=0.0214\}$ we have that $r_\mathrm{s}(z_\mathrm{d})=151.365\; \mathrm{Mpc}$. Our reconstruction procedure is as follows. First, our predefined grammar was constructed on the following  functions: exp, log, polynomials etc. and a set of operations $+,-,\times,\div$, see Table \ref{tab:grammars} for the complete list.

Once the initial population has been constructed, the fitness of each member, which indicates how accurately each individual of the population fits the data, is computed by a $\chi^2$ statistic using the $r_\mathrm{s}(z_\mathrm{d})$ data points directly as input, i.e.
\be\label{eq:chi2}
\chi^2=\sum^N_{i=1}\Big[r_\mathrm{s,i}(z_\mathrm{d})-r_\mathrm{s,GA}(z_\mathrm{d})\Big]^2,
\ee
where $N$ represents the number of data points, which in our case was around 4000, and $r_\mathrm{s}(z_\mathrm{d})_{GA}$ is the fitting function derived by the GA. Notice that in Eq.~(\ref{eq:chi2}) we are not considering uncertainties in each data point since we are taking directly the output value derived with  the code \texttt{CLASS}. 

Then, through a tournament selection process, see Ref.~\cite{Bogdanos:2009ib} for more details, the best-fitting functions in each generation are chosen and the two stochastic operations of crossover and mutation are used. The final output of the code is a mathematical function of $r_\mathrm{s}(z_\mathrm{d})$ that describes the sound horizon at the drag epoch in terms of the various cosmological parameters of interest.

\section{Results\label{sec:results}}
In this section we now present our machine learning fits to the sound horizon at the baryon drag epoch $r_\mathrm{s}(z_\mathrm{d})$. First, we will only include the dependence on the matter and baryon density parameters $\{\Omega_mh^2,\Omega_bh^2\}$, while later we will also consider the effect of massive neutrinos and a varying fine structure constant, i.e. the parameter vectors will be $\{\Omega_mh^2,\Omega_bh^2,\Omega_{\nu}h^2\}$ and $\{\Omega_mh^2,\Omega_bh^2,\alpha/\alpha_0\}$ respectively. 

The computation of the sound horizon is described in Sec.~\ref{sec:GA} and we fit the values with both traditional minimization approaches and with the genetic algorithms. To simplify our notation we make the following definitions that will be used throughout the text: $\omega_b=\Omega_b h^2$, $\omega_m=\Omega_m h^2$ and $\omega_{\nu}=\Omega_{\nu}h^2$. In what follows, we will now describe our approach in more detail and present the results for the various cases.

\subsection{Matter and baryons only}

First, we consider the standard case of matter and baryons, as was also studied in Ref.~\cite{Eisenstein:1997ik} (hereafter denoted as EH). This case was obtained by simulating values for $\Omega_m h^2 \in \left[0.025,  0.5\right]$ and $\Omega_b h^2 \ge 0.0125$ and is given by \cite{Eisenstein:1997ik}
\begin{equation}
r_{\mathrm{s}}\left(z_{\mathrm{d}}\right) \simeq \frac{44.5 \ln \left(\frac{9.83}{\omega_m}\right)}{\sqrt{1+10~\omega_b^{3 / 4}}} \mathrm{Mpc},\label{eq:HE}
\end{equation}
which is accurate up to $\sim 2\%$. Since now the recombination codes have more improved physics (for example an improved post-Saha expansion at early phases of hydrogen recombination, see Refs.~\cite{RubinoMartin:2009ry,Lee:2020obi} for a discussion), we have considered the same parametrization as in EH but with the coefficients as free parameters. By fitting the parametrization to a grid of values for $r_\mathrm{s}(z_\mathrm{d})$ for the range $\Omega_m h^2 \in \left[0.13,  0.15\right]$ and $\Omega_b h^2 \in \left[0.0214, 0.0234\right]$, which is around $10\sigma$ from the Planck best-fit, we find the following improved expression
\begin{equation}
r_{\mathrm{s}}\left(z_{\mathrm{d}}\right) = \frac{45.5337 \ln \left(\frac{7.20376}{\omega_m}\right)}{\sqrt{1+9.98592~\omega_b^{0.801347}}} \;\mathrm{Mpc},\label{eq:HE1}
\end{equation}
which is accurate up to $\sim 0.009\%$. Using the same grid of values with the GA we find the following fit which is even better
\be
r_{\mathrm{s}}(z_{\mathrm{d}}) = \frac{1}{a_1\omega_b^{a_2}+a_3\omega_m^{a_4}+a_5\omega_b^{a_6}\omega_m^{a_7}} \mathrm{Mpc}, \label{eq:HEGA}
\ee
where
\ba
a_1&=& 0.00785436, a_2=0.177084, a_3=0.00912388,\nn\\
a_4&=&0.618711, a_5=11.9611, a_6=2.81343,\nn\\
a_7&=&0.784719.\nn
\ea
In this case, our GA improved  expression given by Eq.~\eqref{eq:HEGA} is accurate up to $\sim 0.003\%$.

Next, we also consider a broader range of values for the parameter grid in order to allow for the fitting function to be used in BAO analyses without compromising its accuracy. In particular, we consider the range $\Omega_m h^2 \in \left[0.05, 0.25\right]$ and $\Omega_b h^2 \in \left[0.016, 0.03\right]$ and we find with the GA the following fit
\be 
r_{\mathrm{s}}(z_{\mathrm{d}}) = \Big[\frac{1}{a_1\omega_b^{a_2}+a_3\omega_b^{a_4}\omega_m^{a_5}+a_6\omega_m^{a_7}}-\frac{a_8}{\omega_m^{a_9}} \Big]\mathrm{Mpc}, \label{eq:HEGA1}
\ee
where
\ba
a_1&=& 0.00257366, a_2=0.05032, a_3=0.013,\nn\\
a_4&=&0.7720642, a_5=0.24346362, a_6=0.00641072,\nn\\
a_7&=& 0.5350899, a_8=32.7525, a_9=0.315473. \nn
\ea
which is accurate up to $\sim 0.018\%$, i.e. a two orders of magnitude improvement from the EH expression of Eq.~\eqref{eq:HE}.

In order to quantify the bias introduced in deriving constraints on the cosmological parameters by using the less accurate expression of Eq.~\eqref{eq:HE}, we will now present the confidence contours and parameter constraints obtained via a Markov chain Monte Carlo (MCMC) with the code \texttt{MontePython 3} of Ref.~\cite{Brinckmann:2018cvx}, using the currently available BAO data as described in Ref.~\cite{Arjona:2020kco} and the aforementioned $r_\mathrm{s}(z_\mathrm{d})$ expressions. As mentioned earlier, $r_\mathrm{s}(z_\mathrm{d})$ and $h\equiv H_0/100$ are degenerate, we in what follows we will consider the combination $r_{s,d}h=r_\mathrm{s}(z_\mathrm{d})h$.

\begin{figure}[!t]
\centering
\includegraphics[width = 0.45\textwidth]{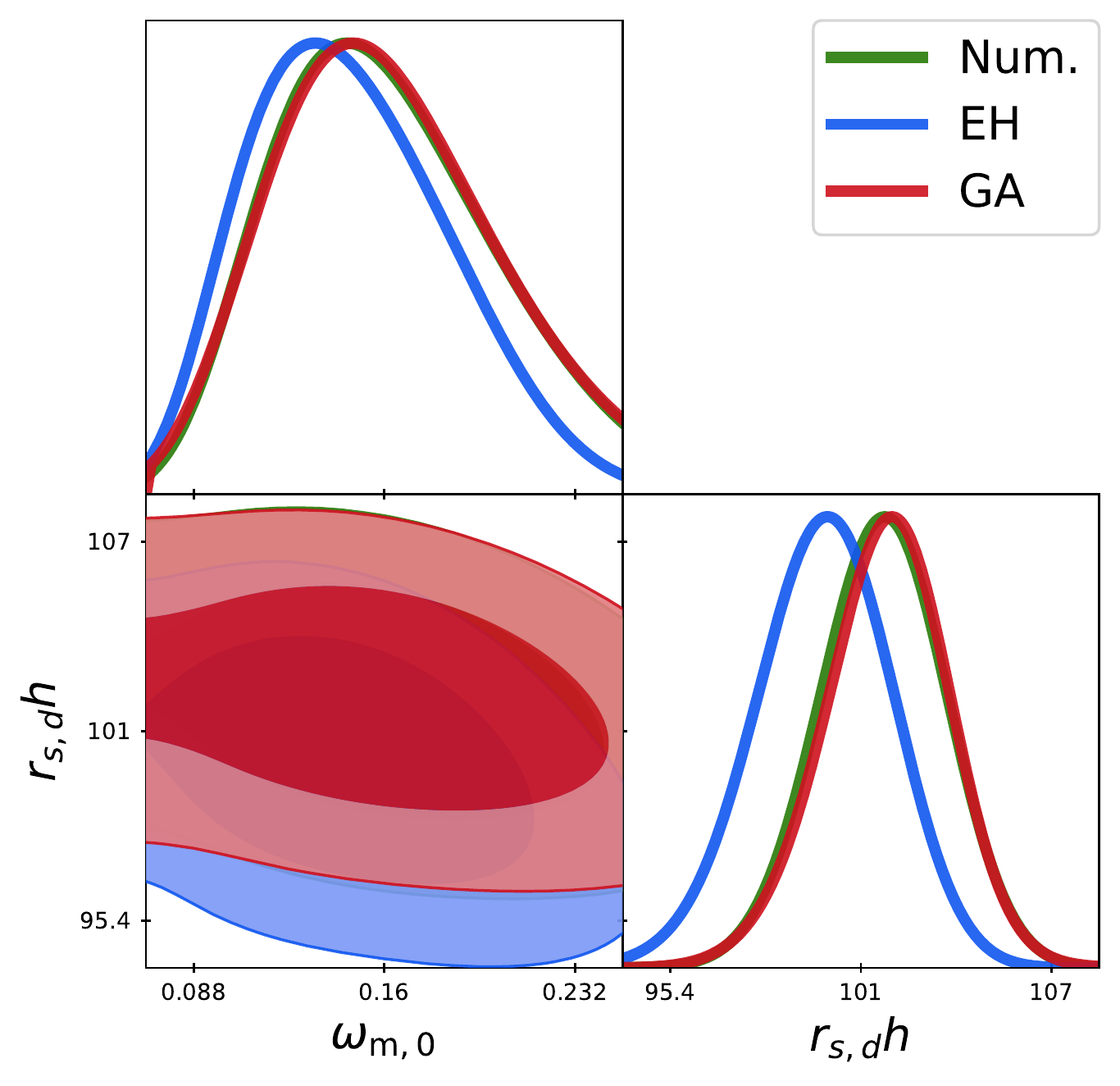}\caption{A comparison of the confidence contours for the expression by Eisenstein-Hu (EH) for the sound horizon given by Eq.~\eqref{eq:HE} (blue contour) against the improved expression found by the machine learning approach (GA) given by Eq.~\eqref{eq:HEGA1} (red contours) and the exact numerical approach (Num.) calculated via \texttt{Hyrec2020} (green contour), using the current BAO data as described in Ref.~\cite{Arjona:2020kco}.\label{fig:ΒΑΟ}}
%\vspace*{-1mm}
\end{figure}

In particular, in Fig.~\ref{fig:ΒΑΟ} we show a comparison of the confidence contours for the EH expression for the sound horizon given by Eq.~\eqref{eq:HE} (blue contour) against the machine learning improved expression (GA) given by Eq.~\eqref{eq:HEGA1} (red contours) and the exact numerical approach (Num.) calculated via \texttt{Hyrec2020} (green contour). Furthermore, in Table \ref{tab:my_label} we show the best-fit, mean and $95\%$ limits for ($\omega_\mathrm{m,0},r_{s,d}h$) obtained from the MCMC runs. 

As can be seen, using the older and less accurate expression biases strongly the constraints for both $\omega_\mathrm{m,0}$ and $r_\mathrm{s}(z_\mathrm{d})h$ by almost half a $\sigma$ and shifts the best-fit $\omega_\mathrm{m,0}$ by $\sim 9.3\%$ from its true value, which is obtained using the full numerical approach. This implies that any analysis, e.g. Refs.~\cite{Beutler:2011hx,Komatsu:2010fb,Bamba:2012cp,Zhai:2018vmm,
Martinelli:2020hud}, using the simple EH formula of Eq.~\eqref{eq:HE} will be biased by about half a $\sigma$ and should be interpreted with some care.

\begin{table}[!t]
    \centering
\begin{tabular}{cccccc} 
 \hline 
Method & Param & best-fit & mean$\pm\sigma$ & 95\% lower & 95\% upper \\ \hline 
Num.& $\omega_\mathrm{m,0}$ &$0.1968$ & $0.1641_{-0.051}^{+0.04}$ & $0.0788$ & $0.251$ \\ 
&$r_{s,d}h$ &$102.1$ & $101.7_{-1.8}^{+1.9}$ & $97.91$ & $105.4$ \\
\hline 
EH & $\omega_\mathrm{m,0}$ &$0.1816$ & $0.1488_{-0.044}^{+0.036}$ & $0.07544$ & $0.2222$ \\ 
& $r_{s,d}h$ &$100.3$ & $99.9_{-1.9}^{+2.2}$ & $95.74$ & $103.9$ \\ 
\hline 
GA & $\omega_\mathrm{m,0}$ &$0.1959$ & $0.1645_{-0.054}^{+0.04}$ & $0.07738$ & $0.2535$ \\ 
& $r_{s,d}h$ &$102.3$ & $101.7_{-1.8}^{+1.9}$ & $97.94$ & $105.5$ \\ 
\hline 
 \end{tabular} \\ 
\caption{The best-fit, mean and $95\%$ limits for ($\omega_\mathrm{m,0},r_{s,d}h$) as discussed in the text. As seen, the older EH approach biases the estimated mean values for the parameters by almost half a $\sigma$, even though they share the same value of the $\chi^2$ at the minimum $\chi^2_\mathrm{min}=10.95$. The contours are shown in Fig.~\ref{fig:ΒΑΟ}.\label{tab:my_label}}
\end{table}

\subsection{Matter, baryons and massive neutrinos}

Next, we also include massive neutrinos and this time we compare with the expression of Ref.~\cite{Aubourg:2014yra}, where the following fit was presented
\begin{equation}
r_\mathrm{s}(z_\mathrm{d}) \approx \frac{55.154 \exp \left[-72.3\left(\omega_{\nu}+0.0006\right)^{2}\right]}{\omega_{m}^{0.25351} \omega_{b}^{0.12807}} \mathrm{Mpc},\label{eq:aubourg}
\end{equation}
which is accurate up to $0.29\%$ within our range of values considered. Notice that this expression is accurate up to $0.021\%$ if we limit to the range within $3\sigma$ of values derived by Planck and that $\omega_{\nu}=0.0107\left(\sum m_{\nu}/1.0\text{eV}\right)$. 

In our case we consider the parameters in the range $\Omega_m h^2 \in \left[0.13,  0.15\right]$, $\Omega_b h^2 \in \left[0.0214 , 0.0234\right]$, which is around $10\sigma$ from Planck, and for the massive neutrinos in the range $0<\sum m_{\nu}<0.6 \text{eV}$. Then, with the GA we find the improved fit which reads as follows
\be\label{eq:neutrino}
r_\mathrm{s}(z_\mathrm{d})=\frac{a_1~e^{a_2\left(a_3+\omega_{\nu}\right)^2}}{a_4~\omega_b^{a_5}+a_6~\omega_m^{a_7}+a_8\left(\omega_b\hspace{1mm}\omega_m\right)^{a_9}}~ \mathrm{Mpc},
\ee
where the coefficients take the following values
\ba
a_1&=& 0.0034917, a_2=-19.972694, a_3=0.000336186,\nn\\
a_4&=&0.0000305, a_5=0.22752, a_6=0.00003142567,\nn\\
a_7&=&0.5453798, a_8=374.14994, a_9=4.022356899,~~~\nn
\ea
which is accurate up to $0.0076\%$, i.e. roughly a factor of three improvement over Eq.~\eqref{eq:aubourg} in the range within $3\sigma$ of Planck and a factor of $\sim 30$ in the broader range.

\subsection{Matter, baryons and the fine structure constant}

Finally, we also consider the effects of a varying fine structure constant on the sound horizon at the drag redshift. The fine structure constant $\alpha$ is already included in the recombination code \texttt{Hyrec2020} \cite{Lee:2020obi,AliHaimoud:2010dx}, thus the only modification in the code that was needed in this case  was passing an extra parameter to \texttt{CLASS}.

Then, we simulate values of the $r_\mathrm{s}(z_\mathrm{d})$ for the range $\Omega_m h^2 \in \left[0.13,  0.15\right]$, $\Omega_b h^2 \in \left[0.0214, 0.0234\right]$ and $\alpha/\alpha_0 \in \left[0.98,1.02\right]$. The range for $\alpha/\alpha_0$ might seem restrictive, but in Ref.~\cite{Ade:2014zfo} it was shown that with current data any variations are constrained to  $\Delta \alpha/\alpha_0\sim 10^{-3}$, while with future large scale structure data and local astrophysical measurements the constraints can be further reduced to $\Delta \alpha/\alpha_0\sim 10^{-6}$. Following the same procedure as before we find the following fitting formula using an EH-like parametrization
\be
r_{\mathrm{s}}\left(z_{\mathrm{d}}\right) = \frac{a_1 \ln \left(\frac{a_2}{\omega_m}\right)}{\sqrt{1+a_3~\omega_b^{a4}}} \big(\alpha/\alpha_0\big)^{a_5}~ \mathrm{Mpc},\label{eq:EHlike}
\ee
which is accurate up to $\sim 0.047\%$ and the parameters are given by  
\ba
a_1&=& 45.417, a_2= 7.15466, a_3= 10.1167,\nn \\
a_4&=& 0.811586,a_5=-1.254537.\nn
\ea 
%Note that in the case of the EH-like parametrization, the exponent of $\alpha/\alpha_0$ is roughly $a_5\simeq -4/3$.

On the other hand, with the GA we have found an improved fit which reads as follows
\be\label{eq:alpha}
r_\mathrm{s}(z_\mathrm{d})=\frac{1}{a_1\omega_b^{a_2}\omega_m^{a_3}\left[\left(\alpha/\alpha_0\right)^{a_4}+\omega_b^{a_5}\omega_m^{a_6}\right]+a_7\omega_m^{a_8}}~ \mathrm{Mpc},
\ee
where the coefficients take the following values
\ba
a_1&=& 0.00730258, a_2=0.088182, a_3=0.099958,\nn\\
a_4&=&1.97913, a_5=0.346626, a_6=0.0092295,\nn\\
a_7&=&0.0074056, a_8=0.8659935,\nn
\ea
which is accurate up to $0.0077\%$ and is roughly a factor of six improvement over the EH-like parametrization of Eq.~\eqref{eq:EHlike}. 

\section{Conclusions\label{sec:conclusions}}
In summary, we have presented extremely accurate machine learning fits to the comoving sound horizon at the baryon drag epoch $r_\mathrm{s}(z_\mathrm{d})$ as a function of cosmological parameters and we compared our results with other expressions found in the literature. In particular, we considered the widely used Eisenstein-Hu fitting formula given by Eq.~\eqref{eq:HE}, which is accurate to the $\sim2\%$ level, and showed how it may strongly bias any constraints on the matter density parameter obtained by using the current BAO data as described in Ref.~\cite{Arjona:2020kco}. 

In particular, we found that the confidence contours are biased by roughly half a sigma, while the matter density parameter $\omega_\mathrm{m,0}$ is shifted at a $\sim9.3\%$ level from its correct value, which is obtained using the full numerical analysis. On the other hand, our machine learning fits given by Eq.~\eqref{eq:HEGA} do not suffer from this issue, as they are accurate to within $\sim 0.003\%$. Furthermore, in our analysis we also considered the effect of massive neutrinos, see Eq.~(\ref{eq:neutrino}) and a varying fine structure constant, see Eq.~(\ref{eq:alpha}), finding that our fits have an improvement of a factor of three to four compared to other simple EH-like parametrizations.

It should be noted though that according to Ref.~\cite{AliHaimoud:2010dx}, \texttt{Hyrec2020} achieves an accuracy of of the order of $\sim10^{-4}$, which is comparable to the precision of the GA results. On the other hand, many forthcoming surveys like Euclid, see Ref.~\cite{EUCLID:2011zbd}, expected to measure the cosmological parameters to about $1\%$ precision, which is about two orders of magnitude larger than the precision of the GA results. As a result, the latter are not expected to bias any analyses with data products from forthcoming surveys in the near term, such as Euclid.

To conclude, we presented machine learning improved expressions for the sound horizon at the drag redshift, which are more accurate in some cases even by two orders of magnitude compared to other similar expressions already found in the literature. The advantage of our approach is that the new expressions do not bias the parameter constraints obtained from BAO data, thus they can be used in BAO analyses coming from current and upcoming surveys to derive cosmological constraints and ease the computational cost that would be required when computing $r_\mathrm{s}(z_\mathrm{d})$ with a full Boltzmann code.

\section*{Acknowledgements}
The authors acknowledge use of the codes:  \texttt{CLASS} version 3.0.1, \texttt{MontePython 3} and the \texttt{HYREC-2} recombination module \texttt{Hyrec2020}. They also acknowledge support from the Research Project  PGC2018-094773-B-C32 and the Centro de Excelencia Severo Ochoa Program SEV-2016-0597. S.~N. also acknowledges support from the Ram\'{o}n y Cajal program through Grant No. RYC-2014-15843. The authors also acknowledge use of the Hydra cluster at the Instituto de F\'isica Te\'orica (IFT), on which some of the numerical computations for this paper took place.

Numerical Analysis Files: The Genetic Algorithm code used by the authors in the analysis of the paper and the expressions of the fits can be found at \href{https://github.com/RubenArjona}{https://github.com/RubenArjona}.

\begin{appendix} %First appendix

\section{Fits for the redshift of the drag epoch and the photon-decoupling surface \label{sec:zd}}

Here we provide some fits for the redshift at the drag epoch $z_\mathrm{d}$, which can be used in Eq.~(\ref{eq:rs_eq}) as a complementary fit instead of the analytic fit of $r_\mathrm{s}(z_\mathrm{d})$ and also a fit to the redshift at the photon-decoupling surface $z_{*}$.

\subsection{The drag redshift $z_\mathrm{d}$}
The fit for the drag redshift from Ref.~\cite{Eisenstein:1997ik} is given by
\begin{equation}
z_{d}=\frac{1291\left(\omega_m\right)^{0.251}}{1+0.659\left(\omega_m\right)^{0.828}}\left[1+b_{1}\left(\omega_b\right)^{b 2}\right],
\end{equation}
where
\ba
b_{1}&=&0.313\left(\omega_m\right)^{-0.419}\left[1+0.607\left(\omega_m\right)^{0.674}\right],\nn\\
b_{2}&=&0.238\left(\omega_m\right)^{0.223},\nn
\ea
and which is accurate up to $\sim 3.7\%$. 

To improve this fit, we simulate values for $z_\mathrm{d}$ in the range $\Omega_m h^2 \in \left[0.13, 0.15\right]$ and $\Omega_b h^2 \in \left[0.0214, 0.0234\right]$ which is around $10\sigma$ from Planck. Then, with the GA we find
\be
z_\mathrm{d}=\frac{1+428.169 \omega_b^{0.256459}\omega_m^{0.616388}+925.56 \omega_m^{0.751615}}{\omega_m^{0.714129}}.
\ee
which is accurate up to $\sim 0.001\%$.

\subsection{The redshift at recombination $z_{*}$}
The fit for the redshift to the photon-decoupling surface $z_{*}$ from Ref.~\cite{Hu:1995en} is given by 
\begin{equation}
z_{*}=1048\left[1+0.00124\left(\Omega_{b} h^{2}\right)^{-0.738}\right]\left[1+g_{1}\left(\Omega_{m} h^{2}\right)^{g_{2}}\right],
\end{equation}
where
\ba
g_{1} &=&\frac{0.0783\left(\Omega_{b} h^{2}\right)^{-0.238}}{1+39.5\left(\Omega_{b} h^{2}\right)^{0.763}}, \nn\\
g_{2} &=&\frac{0.560}{1+21.1\left(\Omega_{b} h^{2}\right)^{1.81}},\nn
\ea
and which is accurate up to $\sim 0.3\%$.

To  improve  this  fit, we  simulate values for $z_{*}$ for the range $\Omega_m h^2 \in \left[0.13, 0.15\right]$ and $\Omega_b h^2 \in \left[0.0214, 0.0234\right]$ which is around $10\sigma$ from Planck. Then, as before, with the GA we find
\be
z_{*}=\frac{391.672\omega_m^{-0.372296}+937.422\omega_b^{-0.97966}}{\omega_m^{-0.0192951}\omega_b^{-0.93681}}+\omega_m^{-0.731631},
\ee
which is accurate up to $\sim 0.0005\%$.

\end{appendix}

%\clearpage
\bibliography{rs}

\end{document}